\documentclass[pra,twocolumn,amsmath,showpacs]{revtex4}
\usepackage{graphicx,bm}
\begin{document}

\author{Hong Ran}
\affiliation{Department of Chemistry, Durham University, South Road,
Durham, DH1~3LE, United Kingdom} \affiliation{Institute of
Theoretical and Computational Chemistry, Key Laboratory of
Mesoscopic Chemistry, School of Chemistry and Chemical Engineering,
Nanjing University, Nanjing 210093, China}
\author{J. Aldegunde\footnote{Present
address: Departamento de Qu{\'\i}mica F{\'\i}sica, Facultad de
Ciencias Qu{\'\i}micas, Universidad de Salamanca, Plaza de los
Ca\'idos s/n, 37008, Salamanca, Spain}}
\email{E-mail: jalde@usal.es}%
\affiliation{Department of Chemistry, Durham University, South
Road, Durham, DH1~3LE, United Kingdom}
\author{Jeremy M. Hutson}
\email{E-mail: J.M.Hutson@durham.ac.uk}%
\affiliation{Department of Chemistry, Durham University, South Road,
Durham, DH1~3LE, United Kingdom}

\title{Hyperfine structure in the microwave spectra\\
of ultracold polar molecules}

\date{\today}

\begin{abstract}
We investigate the microwave spectra of ultracold alkali metal
dimers in magnetic, electric and combined fields, taking
account of the hyperfine structure due to the nuclear spins. We
consider the molecules $^{41}$K$^{87}$Rb and
$^{7}$Li$^{133}$Cs, which are the targets of current
experiments and demonstrate two extremes of large and small
nuclear quadrupole coupling. We calculate the frequencies and
intensities of transitions that may be used to transfer
ultracold molecules between hyperfine states in a magnetic
field, employing different polarizations of microwave
radiation. In an electric field, the hyperfine levels display
narrow avoided crossings at specific fields that we explain in
terms of molecular alignment. The hyperfine splittings that
arise in electric fields may hinder individual addressing in
schemes to use ultracold molecules in quantum computation, but
the structure of the spectra is suppressed in combined fields.
\end{abstract}


%
\maketitle
\section{Introduction}\label{intro}

Ultracold molecules have potential applications in many areas
of physics, ranging from precision measurement to quantum
computing \cite{Carr:NJPintro:2009}. It has recently become
possible to create ultracold alkali metal dimers in their
ground rovibrational state in ultracold atomic gases, by
magnetoassociation followed by state transfer using stimulated
Raman adiabatic passage (STIRAP). This has been achieved for
both polar $^{40}$K$^{87}$Rb \cite{Ni:KRb:2008} and nonpolar
Cs$_2$ \cite{Danzl:ground:2009} and triplet $^{87}$Rb$_{2}$
\cite{Winkler:2007}, and several other systems are being
pursued \cite{Weber:2008, Voigt:2009, Pilch:2009}. It is also
now possible to produce low-lying molecular states by direct
photoassociation \cite{Sage:2005, Hudson:PRL:2008, Viteau:2008,
Deiglmayr:2008, Haimberger:2009}.

Alkali metal dimers have a rich spin structure because of the
presence of two nuclei with non-zero spin. The nuclear spins
interact with one another and with the molecular rotation
through several different terms in the hyperfine Hamiltonian.
In previous work, we have explored the hyperfine structure, and
the effect of applied electric and magnetic fields, for both
polar \cite{Aldegunde:polar:2008} and nonpolar
\cite{Aldegunde:nonpolar:2009} alkali metal dimers.

The microwave spectra of alkali metal dimers are important in a
variety of contexts. They provide opportunities to transfer
polar molecules between spin states and thus to control the
nature of the molecular sample \cite{Ospelkaus:2009}. They are
important in proposals to use ultracold molecules in designs
for quantum computers \cite{DeMille:2002, Andre:2006,
Micheli:2006}, to control intermolecular interactions and
create novel quantum phases \cite{Buechler:2007, Micheli:2007,
Gorshkov:2008}, in condensed-phase models such as the molecular
Hubbard Hamiltonian \cite{Micheli:2006, Wall:2009}, to trap
polar molecules using microwave radiation
\cite{DeMille:trap:2004} and to use cavity-assisted microwave
cooling \cite{Wallquist:2008}. However, work on these topics
has almost invariably neglected the effects of hyperfine
structure.

We have previously given a brief report of the microwave
spectrum of $^{40}$K$^{87}$Rb, taking account of hyperfine
structure and the effects of electric and magnetic fields
\cite{Aldegunde:spectra:2009}. The purpose of the present paper
is to give a more complete account of the spectra and to extend
our predictions to other systems of current experimental
interest. In the present paper we compare and contrast the
predicted spectra for $^{41}$K$^{87}$Rb and $^{7}$Li$^{133}$Cs,
which are the targets of current experiments \cite{Weber:2008,
Deiglmayr:2008} and which demonstrate two extremes of large and
small nuclear quadrupole coupling. We concentrate on spectra
originating in the rotationless ground state, which is the
principal focus of the experimental efforts.

The structure of the paper is as follows. Section
\ref{sec:MolHam} describes the theoretical methods used, while
Sections \ref{sec:hfB}, \ref{sec:hfE} and \ref{sec:hfBE}
discuss the spectra in the presence of magnetic, electric and
combined fields, respectively.

\section{Theoretical methods}
\label{sec:MolHam}

The effective Hamiltonian of a $^{1}\Sigma$ molecule in a
specified vibrational state consists of rotational, hyperfine,
Zeeman and Stark terms \cite{Townes:1975, Brown:2003,
Bryce:2003}. The hyperfine term includes contributions that
represent the nuclear electric quadrupole interaction, the
nuclear spin-rotation interaction, and the scalar and tensorial
nuclear spin-spin interactions. The Zeeman term is dominated by
the nuclear contribution describing the interaction of the
nuclei with the magnetic fields. In the present work we
construct the Hamiltonian as described in ref.\
\onlinecite{Aldegunde:polar:2008} and diagonalize it to obtain
the energy levels and wavefunctions.

The hyperfine coupling constants for $^{41}$K$^{87}$Rb and
$^{7}$Li$^{133}$Cs have been evaluated as described by
Aldegunde {\em et al.}\ \cite{Aldegunde:polar:2008}. The
resulting values are given in Table \ref{tb:hfcc}. The
calculations were carried out at the equilibrium geometries,
$R_{\rm e}=4.07$~\AA\ for KRb \cite{Ross:1990} and $R_{\rm
e}=3.67$~\AA\ for LiCs \cite{Staanum:2007}. They are thus most
suitable for the vibrational ground state but would also be a
reasonable approximation for other deeply bound vibrational
states.

%
%
\begin{table}
\caption{Parameters of the effective Hamiltonians for
$^{41}$K$^{87}$Rb and $^{7}$Li$^{133}$Cs. Rotational constant ($B$),
rotational $g$-factor ($g_{\rm{r}}$), nuclear $g$-factors ($g_{1}$,
$g_{2}$) and nuclear quadrupole ($(eQq)_{1}$, $(eQq)_{2}$),
shielding ($\sigma_{1}$, $\sigma_{2}$), spin-rotation ($c_{1}$,
$c_{2}$), tensor spin-spin ($c_{3}$) and scalar spin-spin ($c_{4}$)
coupling constants. The subscripts 1 and 2 refer to the more
electronegative atom (K or Li) and to the less electronegative one
(Rb or Cs) respectively. See ref.\ \onlinecite{Aldegunde:polar:2008}
for further explanation of the parameters.} \label{tb:hfcc}
\begin{tabular}{ccc}
\hline\noalign{\smallskip} \hline\noalign{\smallskip}
& $^{41}$K$^{87}$Rb & $^{7}$Li$^{133}$Cs \\
\hline\noalign{\smallskip}
$B$ (GHz) & 1.096 & 5.636 \\
$g_{\rm{r}}$ & 0.0138 & 0.0106 \\
$g_{1}$ & 0.143 & 2.171 \\
$g_{2}$ & 1.834 & 0.738 \\
$(eQq)_{1}$ (kHz)& $-$298 & 18.5 \\
$(eQq)_{2}$ (kHz) & $-$1520 & 188 \\
$\sigma_{1}$ (ppm) & 1321 & 108.2\\
$\sigma_{2}$ (ppm) & 3469 & 6242.5 \\
$c_{1}$ (Hz) & 10 & 32 \\
$c_{2}$ (Hz) & 413 & 3014 \\
$c_{3}$ (Hz) & 21 & 140 \\
$c_{4}$ (Hz) & 896 & 1610 \\
\hline\noalign{\smallskip} \hline\noalign{\smallskip}
\end{tabular}
\end{table}

There are three sources of angular momentum in a $^{1}\Sigma$
molecule: the rotational angular momentum ($N$) and the nuclear
spins ($I_{1}$ and $I_{2}$). In the absence of a field, these
couple together to give a total angular momentum $F$. However,
in the presence of electric or magnetic fields, $F$ is
destroyed and the only conserved quantity is the total
projection quantum number $M_{\rm F}$. It is most convenient to
construct the Hamiltonian in an uncoupled basis set  $|NM_{\rm
N}\rangle |I_1M_1\rangle |I_2M_2\rangle$, where $M_{\rm N}$,
$M_1$ and $M_2$ are the projections of $N$, $I_1$ and $I_2$
onto the $z$-axis defined by the direction of the field.

We consider two different polarizations of the microwave field.
When the polarization is parallel to the $z$-axis, the
selection rules for dipole matrix elements between the
uncoupled basis functions are $\Delta N$=$\pm1$, $\Delta
M_{\rm{N}}$=0, $\Delta M_{1}$=0, $\Delta M_{2}$=0 and $\Delta
M_{\rm{F}}$=0. For any other axis of polarization there are
components that are circularly polarized with respect to $z$;
the selection rules for these components are the same except
for $\Delta M_{\rm{N}}$=$\pm 1$ and $\Delta M_{\rm{F}}$=$\pm
1$.

\section{Microwave spectrum in a magnetic field}
\label{sec:hfB}

The experiments that produce ultracold $^{40}$K$^{87}$Rb
\cite{Ni:KRb:2008} and $^{133}$Cs$_{2}$
\cite{Danzl:ground:2009} in their ground rovibrational state
involve two main steps. Pairs of ultracold atoms are first
associated by sweeping the magnetic field across a zero-energy
Feshbach resonance \cite{Hutson:IRPC:2006, Kohler:RMP:2006}.
The resulting Feshbach molecules are then transferred to the
ground rovibrational state by stimulated Raman adiabatic
passage (STIRAP). The molecules remain in a magnetic field. It
is possible in principle to populate a variety of hyperfine
levels. However, the current experiments produce dimers with a
particular value of $M_{\rm{F}}$ determined by the atomic
states used for magnetoassociation, as described below.

%
%
\begin{figure}[t]
\includegraphics[width=0.95\linewidth]{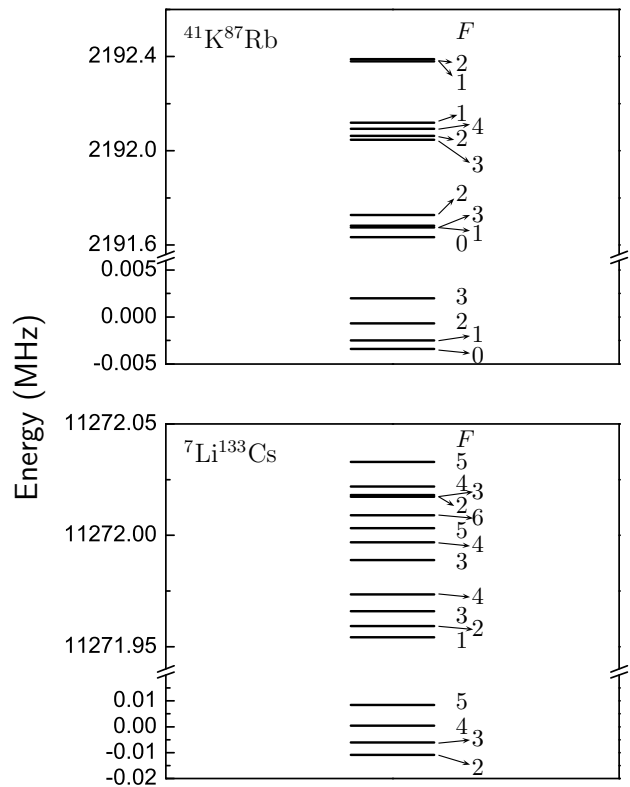}
\caption{\label{fig:01}%
Zero-field hyperfine splitting for $^{41}$K$^{87}$Rb (top
panel) and $^{7}$Li$^{133}$Cs (bottom panel). In each panel,
the levels below the break correspond to $N$=0 and the levels
above the break to $N$=1. The states are labelled by the total
angular momentum quantum number $F$.}
\end{figure}

An important goal is to control the hyperfine state in which
molecules are produced. In particular, the absolute ground
state is stable to inelastic collisions with other ground-state
species. This section explores how microwave transitions can be
used to transfer molecules between hyperfine states, as has now
been achieved experimentally for $^{40}$K$^{87}$Rb
\cite{Ospelkaus:2009}. We consider the use of (i) $z$-polarized
microwave radiation to transfer population between $N=0$ states
of the same $M_{\rm F}$ and (ii) radiation with a circularly
polarized component to transfer population into the ground
hyperfine state by changing $M_{\rm{F}}$. We discuss the
mechanism of the transfer and the experimental conditions that
favor or hinder it.

The zero-field energy level patterns for the $N$=0 and $N$=1
levels of $^{41}$K$^{87}$Rb and $^{7}$Li$^{133}$Cs are shown in
Fig.~\ref{fig:01}. The levels are labeled by the total angular
momentum $F$, obtained by coupling the rotational angular
momentum $N$ and the nuclear spins $I_{1}$ and $I_{2}$. The
nuclear spin is 3/2 for $^{41}$K, $^{87}$Rb and $^{7}$Li and
7/2 for $^{133}$Cs. The splitting of the $N$=0 states is due to
the scalar spin-spin interaction and amounts to just a few kHz.
The splitting of the $N$=1 levels is caused mainly by the
nuclear quadrupole interaction and amounts to 800~kHz for
$^{41}$K$^{87}$Rb and 100~kHz for $^{7}$Li$^{133}$Cs. The
difference stems from the very small nuclear quadrupole
coupling constants for the $^{7}$Li$^{133}$Cs molecule,
attributable to the particularly small values of the electric
quadrupoles for the $^{7}$Li and $^{133}$Cs nuclei
\cite{Stone:2005}.

\begin{figure}[t]
\includegraphics[width=0.95\linewidth]{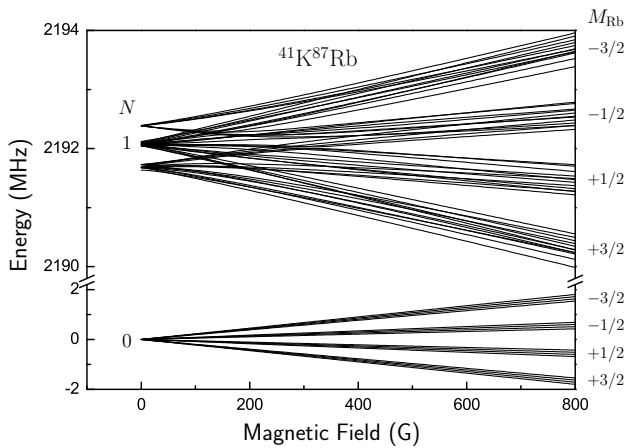}
\caption{\label{fig:02}%
Zeeman splitting of the $^{41}$K$^{87}$Rb hyperfine
levels for $N$=0 and $N$=1.}
\end{figure}

Each of the zero-field levels in Fig.~\ref{fig:01} is
(2$F$+1)-fold degenerate. This degeneracy can be lifted by
applying a magnetic field. The energy pattern can be quite
complicated as it consists of $(2N+1)(2I_{1}+1)(2I_{2}+1)$
states for each rotational level $N$. Fig.\ \ref{fig:02} shows
the Zeeman splitting for the $N$=0 and 1 hyperfine levels of
$^{41}$K$^{87}$Rb and includes all the 16 $N$=0 and the 48
$N$=1 hyperfine states. For $^{7}$Li$^{133}$Cs the number of
levels is even larger and they will not be shown.

%
%

%
%
\begin{figure}[t]
\includegraphics[width=0.95\linewidth]{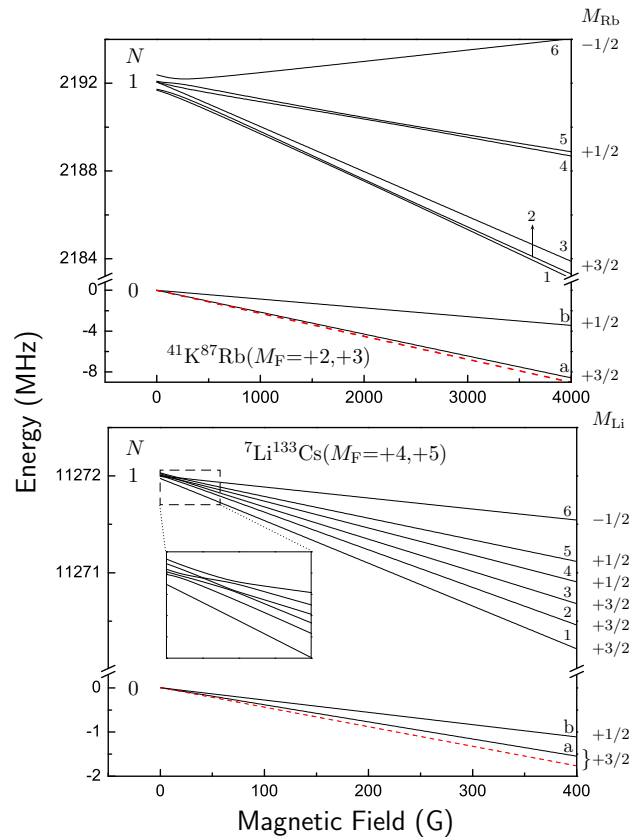}
\caption{\label{fig:03}%
Zeeman splitting of the hyperfine levels for $^{41}$K$^{87}$Rb
($M_{\rm{F}}$=+2) and $^{7}$Li$^{133}$Cs ($M_{\rm{F}}$=+4).
Labels a-b and 1-6 identify the $N$=0 and $N$=1 states
respectively. The small panel shows an expanded view of the
region of avoided crossings for the $N$=1 levels of
$^{7}$Li$^{133}$Cs. The $^{41}$K$^{87}$Rb ($N$=0,
$M_{\rm{F}}$=+3) and $^{7}$Li$^{133}$Cs ($N$=0,
$M_{\rm{F}}$=+5) levels, which are the absolute ground states
except at low magnetic fields, are also included (red dashed
lines).}
\end{figure}

%
%
\begin{figure*}[t]
\includegraphics[width=0.95\linewidth]{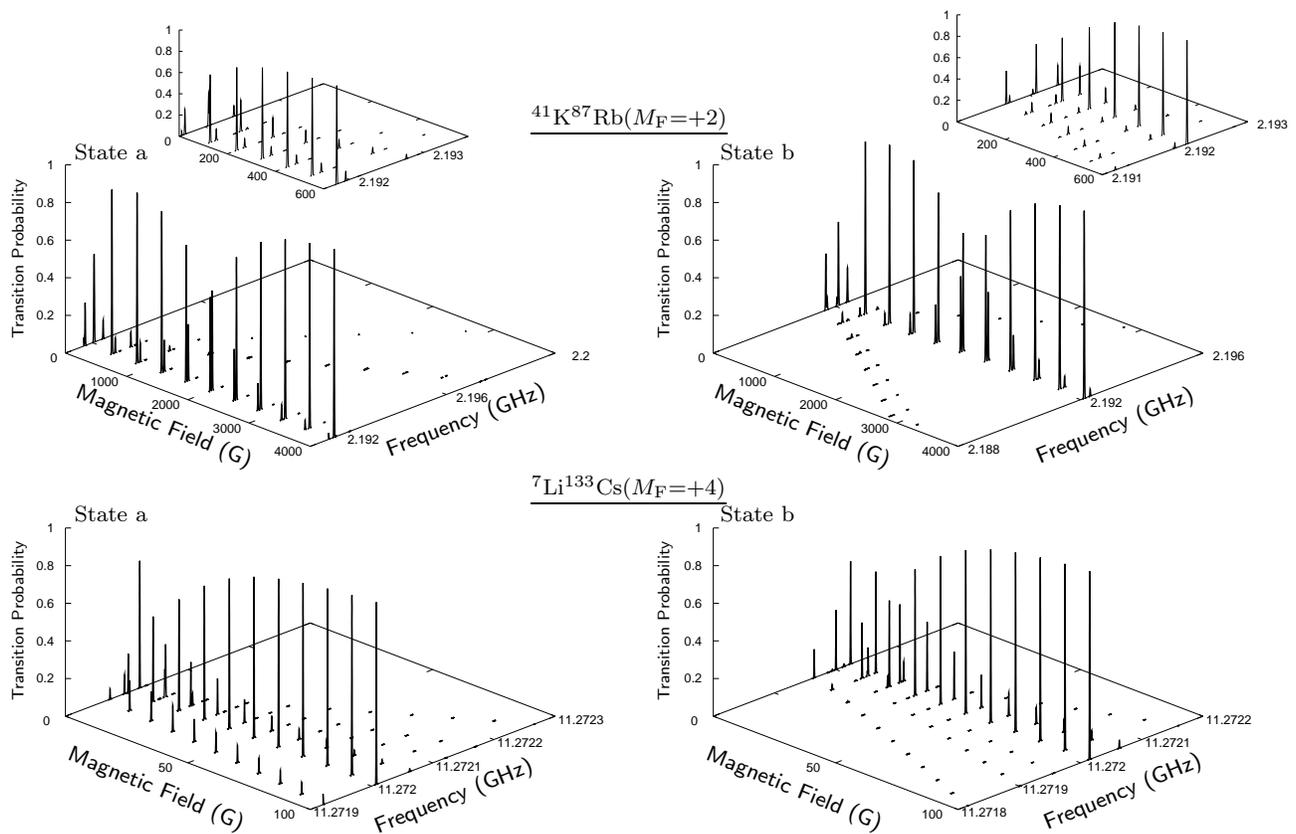}
\caption{\label{fig:04}%
Relative intensity for the microwave transitions from the two
$N$=0 hyperfine states (levels a and b) to the $N$=1 hyperfine
states (levels 1-6) for $^{41}$K$^{87}$Rb $(M_{\rm F}$=+2) (top
panels) and $^{7}$Li$^{133}$Cs $(M_{\rm F}$=+4) (bottom
panels). Enlargements of the 0-600~G region for
$^{41}$K$^{87}$Rb are shown. The microwave radiation is
polarized parallel to the magnetic field. The most intense
transition for each molecule is assigned a peak intensity of
1.}
\end{figure*}

We will first analyze transitions driven by a microwave field
polarized parallel to the $z$-axis, which conserve the
projection of the total angular momentum $M_{\rm{F}}$.
Fig.~\ref{fig:03} shows the Zeeman splitting and avoided
crossings of the $^{41}$K$^{87}$Rb ($M_{\rm{F}}$=+2) and
$^{7}$Li$^{133}$Cs ($M_{\rm{F}}$=+4) hyperfine levels. The
reason we choose these $M_{\rm F}$ values is that
magnetoassociation conserves $M_{\rm F}$ and is usually carried
out using atoms in their absolute ground state in the magnetic
field, which is ($f$=$I$$-$1/2,$m_{\rm f}$=$I$$-$1/2) for all
alkali metal atoms except $^{40}$K. This is ($f$=1,$m_{\rm
f}$=+1) for $^{7}$Li, $^{41}$K, $^{87}$Rb and ($f$=3,$m_{\rm
f}$=+3) for $^{133}$Cs. Feshbach molecules are thus likely to
be formed initially in $M_{\rm{F}}$=+2 and +4 for
$^{41}$K$^{87}$Rb and $^{7}$Li$^{133}$Cs respectively. In
current experimental configurations, the STIRAP process used to
transfer the molecules to the rovibrational ground state also
conserve $M_{\rm{F}}$. The hyperfine state formed in this way
is not generally the absolute ground state, which has $M_{\rm
F}$=$I_1$+$I_2$.

There are important differences between dimers that contain
$^{40}$K and those that do not. $^{40}$K is the only stable
isotope of any alkali metal to have a negative $g$-factor and
an inverted hyperfine structure. Because of this, the atomic
ground state in a magnetic field is ($f$=$I_{\rm
K}$+1/2=9/2,$m_{\rm f}$=$-$9/2) and heteronuclear Feshbach
molecules formed from ground-state atoms have $M_{\rm
F}$=$I_1$$-$5 \cite{Ospelkaus:2008}. The lowest molecular
hyperfine state at high magnetic field has $M_{\rm
F}$=$I_1$$-$$I_{\rm K}$ because of the negative value of
$g_{\rm K}$ for $^{40}$K.

As the magnetic field increases from zero, the levels of each
$F$ split into components labeled by $M_{\rm F}$, as shown in
Figs.\ \ref{fig:02} and \ref{fig:03}. Avoided crossings occur
between states with the same value of $M_{\rm{F}}$. While for
the $N$=0 levels the crossings occur at small magnetic fields
proportional to $|c_{4}/(g_{1}-g_{2})|$, for the $N$=1 states
they take place at larger fields because of the larger
zero-field splittings. Above the crossings, $F$ is completely
destroyed. For both molecules the $N$=0 avoided crossings occur
at fields below 20~G and are hard to see in Figs.~\ref{fig:02}
and \ref{fig:03}. The $N$=1 crossings take place at fields that
depend on zero-field splittings $E_0$ divided by factors
involving the two nuclear $g$-factors. Because of the very
small zero-field splittings between some pairs of states, some
of the avoided crossings can occur at fields as low as a few
Gauss, though others are at much higher fields.

The nuclear Zeeman term, which dominates the Hamiltonian at
high field, is diagonal in the uncoupled basis set. Because of
this, the individual projections of the nuclear spins $M_{1}$
and $M_{2}$ become nearly good quantum numbers at sufficiently
high field. When the $g$-factors for the two nuclei are
substantially different, as for $^{41}$K$^{87}$Rb, the energy
levels for each rotational state separate into distinct groups
as the field increases (see Fig.~\ref{fig:02}). The $g$-factor
for $^{87}$Rb is much larger than that for $^{41}$K, so that
$M_{\rm{Rb}}$ becomes a nearly good quantum number at much
smaller fields than $M_{\rm{K}}$ and the states gather together
in groups characterized by a well defined value of
$M_{\rm{Rb}}$: the lowest set corresponds to $M_{\rm{Rb}}$=+3/2
and the top set to $M_{\rm{Rb}}$=$-3/2$. Similar behavior was
observed for $^{40}$K$^{87}$Rb in ref.\
\onlinecite{Aldegunde:spectra:2009}.

For $^{41}$K$^{87}$Rb with $M_{\rm F}$=+2 the levels within
each group display shallow avoided crossings and $M_{\rm{K}}$
does not become well defined until fields above about 3000~G.
For $^7$Li$^{133}$Cs, by contrast, the avoided crossings are
complete by 40~G. In general, the number of levels in each
group is $N$+$I_-$+$1$$-$$|M_{\rm F}$$-$$M_+|$, where $I_\pm$
and $M_\pm$ are quantum numbers for the nucleus with the
larger/smaller $g$-factor. When there is more than 1 level in a
group there may be high-field crossings between them at a field
of approximately $E_0/g_-\mu_{\rm N}$, where $g_-$ is the
smaller of the two $g$-factors and $\mu_{\rm N}$ is the nuclear
magneton. For $^{41}$K$^{87}$Rb ($N$=1), setting $E_0$ to the
full range of zero-field energies gives an upper bound of about
5500 G, though the highest crossings in Fig.\ \ref{fig:03}
actually occur around 2000~G. For $^7$Li$^{133}$Cs the
corresponding upper bound is about 140~G.

We now consider the microwave transitions that are available to
transfer molecules between hyperfine states with the same value
of $M_{\rm F}$. Fig.\ \ref{fig:04} shows the spectra for
transitions between the two $N$=0 hyperfine states (levels a
and b) and their $N$=1 counterparts (levels 1-6) for
$^{41}$K$^{87}$Rb ($M_{\rm{F}}$=+2) and $^{7}$Li$^{133}$Cs
($M_{\rm{F}}$=+4) as a function of the magnetic field. The
microwave field is $z$-polarized. Both molecules show a
multi-line spectrum at low field but the spectra become
dominated by a single transition in the strong-field limit.
However, while a 100~G field is enough to cause this change for
$^{7}$Li$^{133}$Cs, more than 4000~G is necessary for
$^{41}$K$^{87}$Rb.

The progression from multi-line spectra to
single-line-dominated spectra arises because the selection
rules for $z$-polarized light only permit transitions
\begin{equation}\label{eq:selrul}
|N=0,M_{\rm{N}},M_{1},M_{2}\rangle \leftrightarrow
|N=1,M_{\rm{N}},M_{1},M_{2}\rangle
\end{equation}
where the angular momentum projections do not change. At low
magnetic fields the $|N,M_{\rm{N}},M_{1},M_{2}\rangle$
functions are ``spread" over different eigenstates and multiple
transitions from each $N$=0 state appear. As the magnetic field
increases the eigenstates are better approximated by the
$|N,M_{\rm{N}},M_{1},M_{2}\rangle$ basis functions. This takes
place at much lower magnetic fields for $N$=0 than for $N$=1.
In the strong-field limit the only permitted transition from
each $N$=0 hyperfine state is given by the selection rule
(\ref{eq:selrul}). This occurs at lower field for
$^{7}$Li$^{133}$Cs than for $^{41}$K$^{87}$Rb because $M_{\rm
K}$ is not a good quantum number until very high field in the
latter case.

These spectra are important because they clarify under what
conditions the $N$=0$\leftrightarrow$$N$=1 microwave
transitions can be used to transfer population between
different $N$=0 hyperfine states corresponding to the same
$M_{\rm{F}}$ through a two-photon process that uses the $N$=1
levels as intermediate states. This will be feasible if there
is at least one $N$=1 level that displays a significant
intensity from both the $N$=0 states involved in the transfer.
This requires substantially smaller magnetic fields for
$^{7}$Li$^{133}$Cs than for $^{41}$K$^{87}$Rb.

%
%
\begin{figure}[t]
\includegraphics[width=0.95\linewidth]{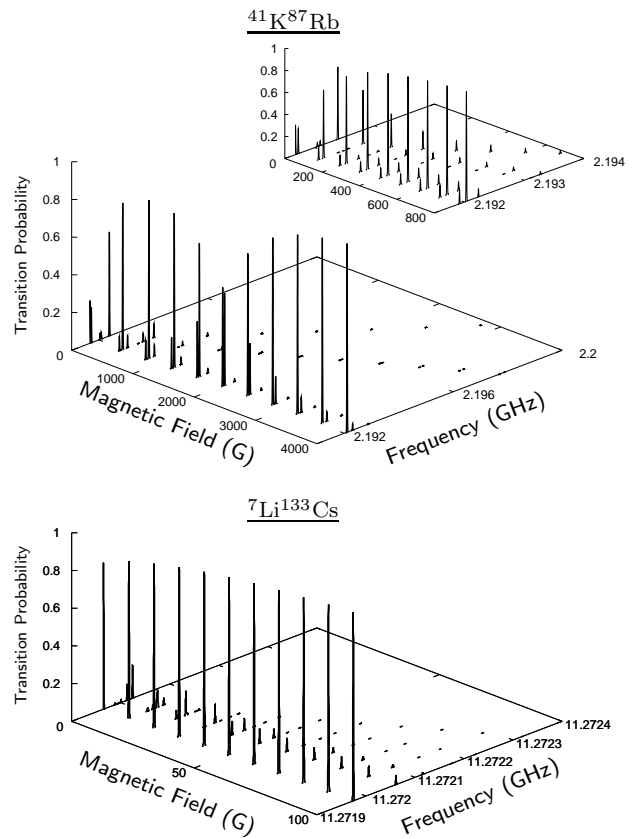}
\caption{\label{fig:05}%
Relative intensities for the transitions between the ($N$=0,
$M_{\rm{F}}$=+3) and ($N$=1, $M_{\rm{F}}$=+2) states of
$^{41}$K$^{87}$Rb (top panels) and between the ($N$=0,
$M_{\rm{F}}$=+5) and ($N$=1, $M_{\rm{F}}$=4) states of
$^{7}$Li$^{133}$Cs (bottom panel). The microwave radiation
is circularly polarized. The most intense transition
for each molecule is assigned a peak intensity of 1.}
\end{figure}

%
%
\begin{figure}[t]
\includegraphics[width=0.95\linewidth]{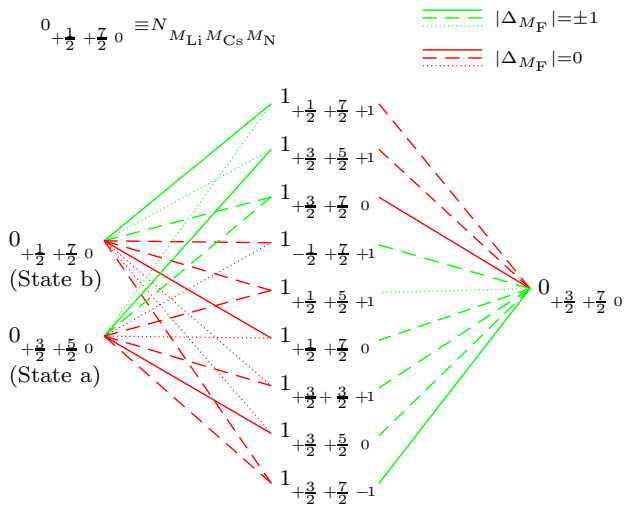}
\caption{\label{fig:06}%
Propensity rules for the transitions available to transfer
$^{7}$Li$^{133}$Cs between the ($N$=0, $M_{\rm{F}}$=+4) states
(a and b, left side) and the absolute ground state with
$M_{\rm{F}}$=+5 (right side) via $N$=1 levels with
$M_{\rm{F}}$=+4 and +5 (center). In the left column the
energy levels are ordered according to their energy in the
strong-field region and in the central column the 6 (3)
$M_{\rm{F}}$=+4 (5) $N$=1 levels are shown separately at the
bottom (top) and then ordered by the same criterion. Green
(red) lines indicate transitions driven by circularly ($z$)
polarized light. Solid, dashed and dotted lines represent
transitions where none, one or two nuclear spin projections
change and correspond to strong, medium and weak lines in the
spectra.}
\end{figure}

Microwave radiation whose polarization is not parallel to the
external field can be used to transfer population between
hyperfine states with different $M_{\rm{F}}$. For
$^{41}$K$^{87}$Rb and $^{7}$Li$^{133}$Cs the absolute ground
state corresponds to $M_{\rm{F}}$=+3 and +5 respectively except
at very low magnetic field (see Fig.~\ref{fig:03}). It is
therefore possible to reach the absolute ground state from
$^{41}$K$^{87}$Rb ($M_{\rm{F}}$=+2) and $^{7}$Li$^{133}$Cs
($M_{\rm{F}}$=+4) if one of the two photons is circularly
polarized. Whether this photon is responsible for the first or
the second transition gives rise to two possible experimental
implementations.

Fig.\ \ref{fig:05} shows the spectra for $M_{\rm F}$-changing
transitions between the absolute ground states of $^{41}$K$^{87}$Rb
and $^{7}$Li$^{133}$Cs and the $N$=1 levels shown in Figs.\
\ref{fig:03} and \ref{fig:04}. In both cases the microwave field is
circularly polarized. As for $\Delta M_{\rm F}$=0 transitions, there
are several lines with significant intensity at low field but a
single line dominates at sufficiently high field. The changeover
occurs at considerably higher fields for $^{41}$K$^{87}$Rb than for
$^{7}$Li$^{133}$Cs.

The propensity rules for the possible paths are summarized in
Fig.~\ref{fig:06} for $^{7}$Li$^{133}$Cs. Solid, dashed and
dotted lines represent transitions where none, one or two
nuclear spins projections change and correlate with strong,
medium and weak lines in the spectra. For either initial state
(a or b), there are several paths to the lowest hyperfine state
that combine one strong transition (solid line) with one weaker
transition (dashed line). However, it should be noted that the
propensity rules become stronger (more like selection rules) as
the magnetic field increases. The overall two-photon transfer
will therefore be more difficult at high fields and forbidden
in the strong-field limit.

The propensity rules are very similar for $^{41}$K$^{87}$Rb
($N$=0,$M_{\rm{F}}$=+2)$\leftrightarrow$($N$=0,$M_{\rm{F}}$=+3).
The number and ordering of the levels and the distribution of
strong, medium and weak transitions are identical to those for
$^{7}$Li$^{133}$Cs. However, in $^{41}$K$^{87}$Rb, $M_{\rm K}$
is not a nearly good quantum number until very high field so
the propensity rules are not as strong.

In summary, the transfer of population between different $N$=0
hyperfine states using microwave transitions will be
facilitated by a larger mixture between the
$|N$=$1,M_{\rm{N}},M_{1},M_{2}\rangle$ basis functions, which
can be achieved by carrying out the transfer at low magnetic
fields. The actual fields required depend on the magnitudes of
the nuclear quadrupole coupling constants and the nuclear
$g$-factors. The transfer is not possible in the limiting
strong-field region.

%
%
\begin{figure}[t]
\includegraphics[width=0.85\linewidth]{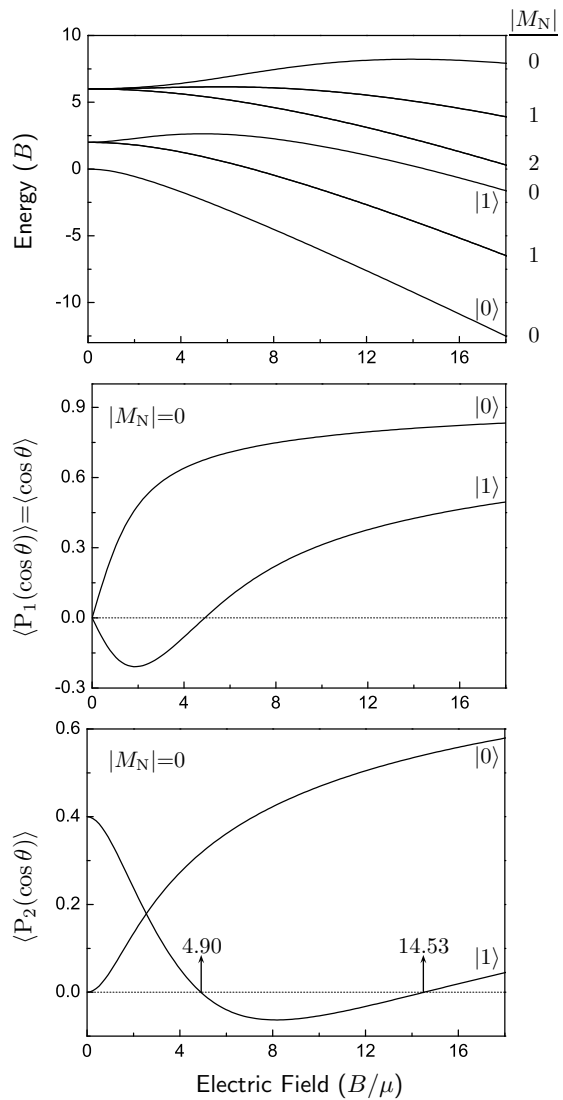}
\caption{\label{fig:07}%
Stark energy levels for a $^{1}\Sigma$ diatomic molecule
correlating with $N$=0 to 2 (top panel). The labels $|0\rangle$
and $|1\rangle$ identify the levels used as qubits. The
hyperfine structure is not included. The central and bottom
panels show the expectation values of the Legendre polynomials
$P_{1}(\cos\theta)$ and $P_{2}(\cos\theta)$ for $|0\rangle$ and
$|1\rangle$.}
\end{figure}

\section{Microwave spectrum in an electric field} \label{sec:hfE}

The microwave spectra of ultracold molecules in electric fields
are important in many contexts, ranging from quantum
information processing \cite{DeMille:2002, Andre:2006,
Micheli:2006} to the creation of novel quantum phases
\cite{Buechler:2007, Micheli:2007}. In this section we explore
the effect of hyperfine structure on these spectra, focusing
particularly on the relevance for implementation of a quantum
computer.

The features of the spectrum in the presence of an electric
field will differ drastically from those found for a magnetic
field. Electric fields do not conserve parity and cause strong
mixing of rotational levels. The strength of the Stark effect
depends on the molecular electric dipole moment $\mu$, 0.57~D
for KRb \cite{Ni:KRb:2008} and 5.5~D for LiCs
\cite{Aymar:2005}.

DeMille has proposed a quantum computing implementation in
which the qubits are ultracold alkali dimers oriented along
($|0\rangle$) or against ($|1\rangle$) an external electric
field \cite{DeMille:2002}. The field strength changes with the
position in the device so as to allow individual addressing of
the molecules, which is performed using microwave transitions.
The $|0\rangle$ and $|1\rangle$ states are mixtures of the
$M_{\rm{N}}$=0 levels that correlate with $N$=0 and 1
respectively at zero field. The goal of this section is to
ascertain whether the hyperfine structure will interfere with
the addressing process.

%
%
\begin{figure*}[t]
\includegraphics[width=0.93\linewidth]{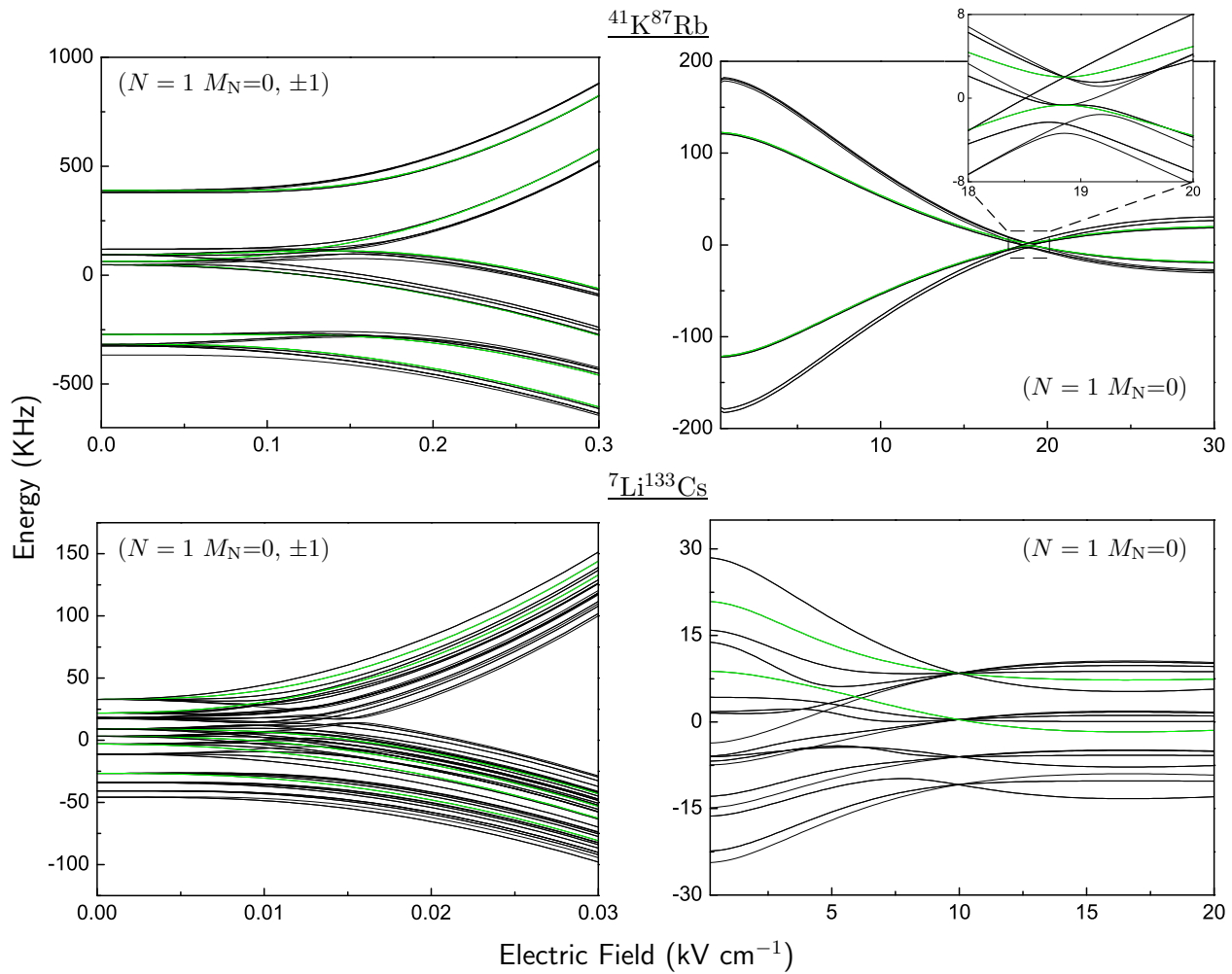}
\caption{\label{fig:08}%
Stark splitting of the hyperfine levels correlating with $N$=1
for $^{41}$K$^{87}$Rb (upper panels) and $^{7}$Li$^{133}$Cs
(lower panels). The left-hand panels include all the hyperfine
states and show the low-field region where the electric field
separates the $|M_{\rm{N}}|$=0 (upper branch) and 1 (lower
branch) levels. The right-hand panels show the splitting for
the $M_{\rm{N}}$=0 states. The levels in each panel are shown
relative to their field-dependent average energy and those
corresponding to $^{41}$K$^{87}$Rb ($M_{\rm{F}}$=+2) and
$^{7}$Li$^{133}$Cs ($M_{\rm{F}}$=+4) are shown in green.}
\end{figure*}

The top panel of Fig.~\ref{fig:07} shows the lowest Stark
energy levels of a $^{1}\Sigma$ diatomic molecule, neglecting
nuclear spin, as a function of the reduced field in units of
$B/\mu$, where $B$ is the rotational constant of the molecule.
The electric field orients the molecule and produces a
space-fixed dipole moment $d=\mu\langle\cos\theta\rangle$,
where $\theta$ is the angle between the internuclear axis and
the electric field. The central panel in Fig.~\ref{fig:07}
shows the degree of {\em orientation}, defined as the
expectation value of $\cos\theta$, for the states $|0\rangle$
and $|1\rangle$. For electric fields in the interval
(2-5)$B/\mu$, the operating range of fields in DeMille's
proposal, the $|0\rangle$ ($|1\rangle$) state corresponds to an
orientation of the permanent dipole parallel (antiparallel) to
the direction of the field. In the strong-field limit, the
expectation value approaches unity for both molecular states,
corresponding to a perfectly parallel orientation of the
dipole.

As in Section \ref{sec:hfB}, we begin by analyzing the behavior
of the energy levels. Because of the choice for $|0\rangle$ and
$|1\rangle$, we focus on $M_{\rm{N}}$=0 states and transitions
driven by a $z$-polarized microwave field.

Fig.\ \ref{fig:08} shows the Stark splitting of the hyperfine
levels correlating with $N$=1 for $^{41}$K$^{87}$Rb and
$^{7}$Li$^{133}$Cs. The field-dependent average energy has been
subtracted so that the hyperfine structure can be appreciated.
Small electric fields, 0.3 kV/cm for $^{41}$K$^{87}$Rb and 0.02
kV/cm for $^{7}$Li$^{133}$Cs, are enough to make $M_{\rm{N}}$ a
nearly good quantum number and to separate the $|M_{\rm{N}}|$=0
and 1 energy levels.

The $M_{\rm{N}}$=0 states display a striking feature as a
function of the electric field. At a field of 18.8 kV/cm for
$^{41}$K$^{87}$Rb or 9.95 kV/cm for $^{7}$Li$^{133}$Cs, all the
levels come very close together and almost cross. At this point
the hyperfine splitting reduces to that caused by the (very
small) scalar spin-spin interaction, as shown in the inset for
$^{41}$K$^{87}$Rb. The effects of the nuclear quadrupole, the
scalar spin-spin and the nuclear spin-rotation interactions
vanish and the splitting simplifies and coincides with that
found for the $N$=0 levels in the absence of fields. At this
point the total nuclear spin obtained by coupling $I_1$ and
$I_2$ is a good quantum number. In addition to their effect on
the microwave spectra, such close avoided crossings may cause
nonadiabatic transitions for molecules in time-varying electric
fields.

%
%
\begin{figure*}[t]
\includegraphics[width=0.95\linewidth]{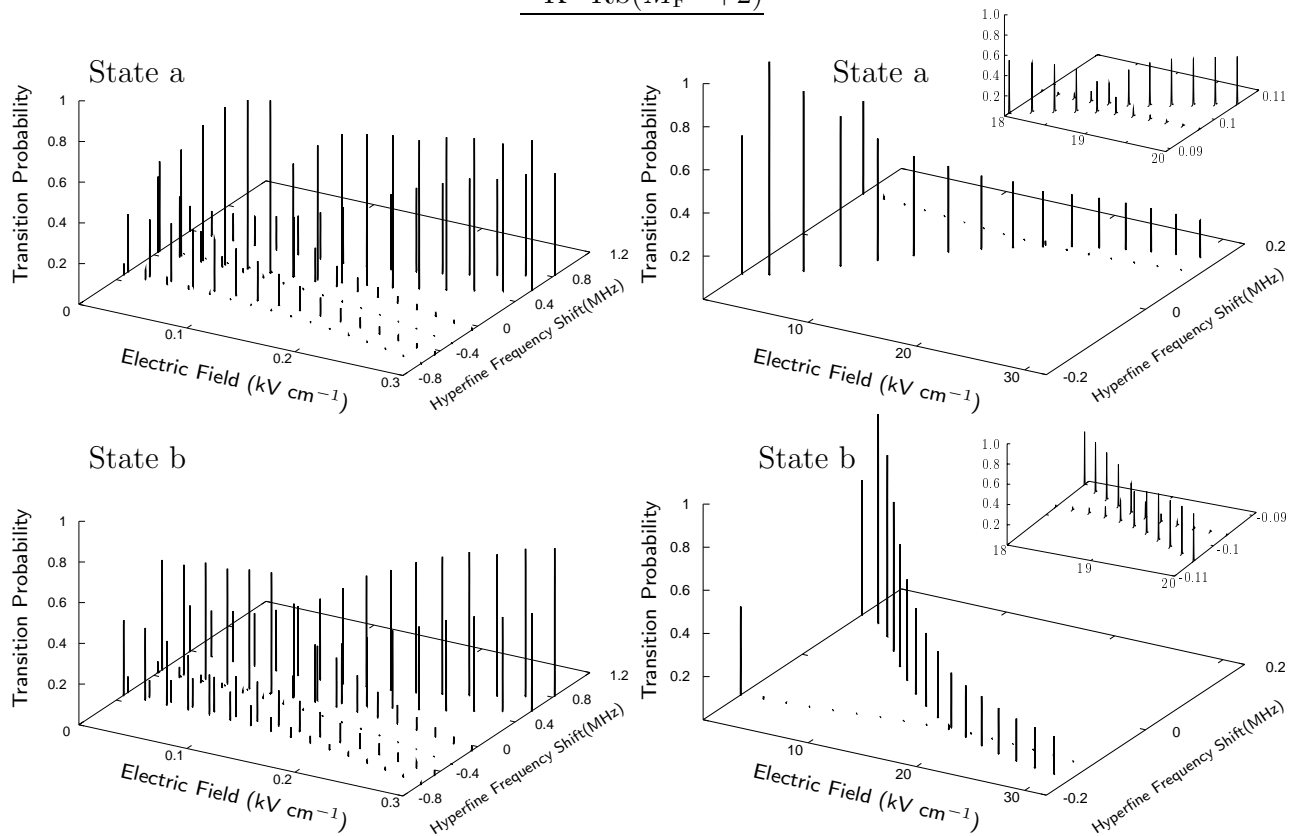}
\caption{\label{fig:09}%
Relative intensities for the transitions between the levels
correlating with $N$=0 (states a and b) and $N$=1 for
$^{41}$K$^{87}$Rb ($M_{\rm{F}}$=+2). The microwave field is
polarized parallel to the electric field. The left-hand panels
show the spectra for low electric fields where $M_{\rm{N}}$ is
not yet a good quantum number. The right-hand panels show
spectra at higher fields where $M_{\rm{N}}$ is well defined and
only transitions to the ($N$=1, $M_{\rm{N}}$=0) branch are
permitted. In order to keep the spectrum on a single frequency
scale as a function of the electric field, we plot the
hyperfine frequency shift instead of the absolute frequency.
This is obtained by subtracting the frequency for the
transition in the absence of hyperfine splittings. The most
intense transition is assigned a peak intensity of 1.}
\end{figure*}

Near-crossings like those shown in Fig.~\ref{fig:08} will be a
universal feature for polar molecules in $^1\Sigma$ states. The
nuclear quadrupole interaction depends on the electric field
gradient created by the electrons at the positions of the
nuclei. This is a cylindrically symmetric second-rank tensor
$V^{(2)}_q$ with only a $q=0$ component in the molecule-fixed
frame. However, the nuclei experience the electric field
gradient in the space-fixed frame, with components $V^{(2)}_p$
given by
\begin{equation}
V^{(2)}_p = D^2_{p0}(\phi,\theta,0) V^{(2)}_{q=0},
\end{equation}
where $D^2_{p0}(\phi,\theta,0)$ is a Wigner rotation matrix and
$\phi$ is the azimuthal angle. For the $M_{\rm N}$=0 levels,
only the component with $p=0$ has diagonal matrix elements, and
$D^2_{00}(\phi,\theta,0)=P_2(\cos\theta)$, the second Legendre
polynomial. The time-averaged field gradient is thus
proportional to the degree of {\em alignment}, defined as
$\langle P_2(\cos\theta)\rangle$. As shown in the bottom panel
of Fig.~\ref{fig:07}, this passes through zero for state
$|1\rangle$ at two specific values of the electric field. The
first zero occurs at an electric field equal to 4.90
$B/\mathrm{\mu}$ and coincides with the points where the
hyperfine structure simplifies in Fig.~\ref{fig:08}. The second
zero appears at larger fields (14.53 $B/\mathrm{\mu}$) that are
outside the range plotted in Fig.~\ref{fig:08}. The tensorial
spin-spin interaction is also proportional to $\langle
P_2(\cos\theta)\rangle$, so vanishes at the same point, while
the nuclear spin-rotation interaction has no diagonal matrix
elements for $M_{\rm{N}}$=0 at any value of the field.

For the $|0\rangle$ state, which correlates with $N$=0,
$\langle P_2(\cos\theta)\rangle$ increases monotonically with
field, from zero at zero field to unity in the high-field
limit. The nuclear quadrupole and tensorial spin-spin
splittings thus also increase monotonically and there are no
close avoided crossings \cite{Aldegunde:polar:2008}.

%
%
\begin{figure*}[t]
\includegraphics[width=0.95\linewidth]{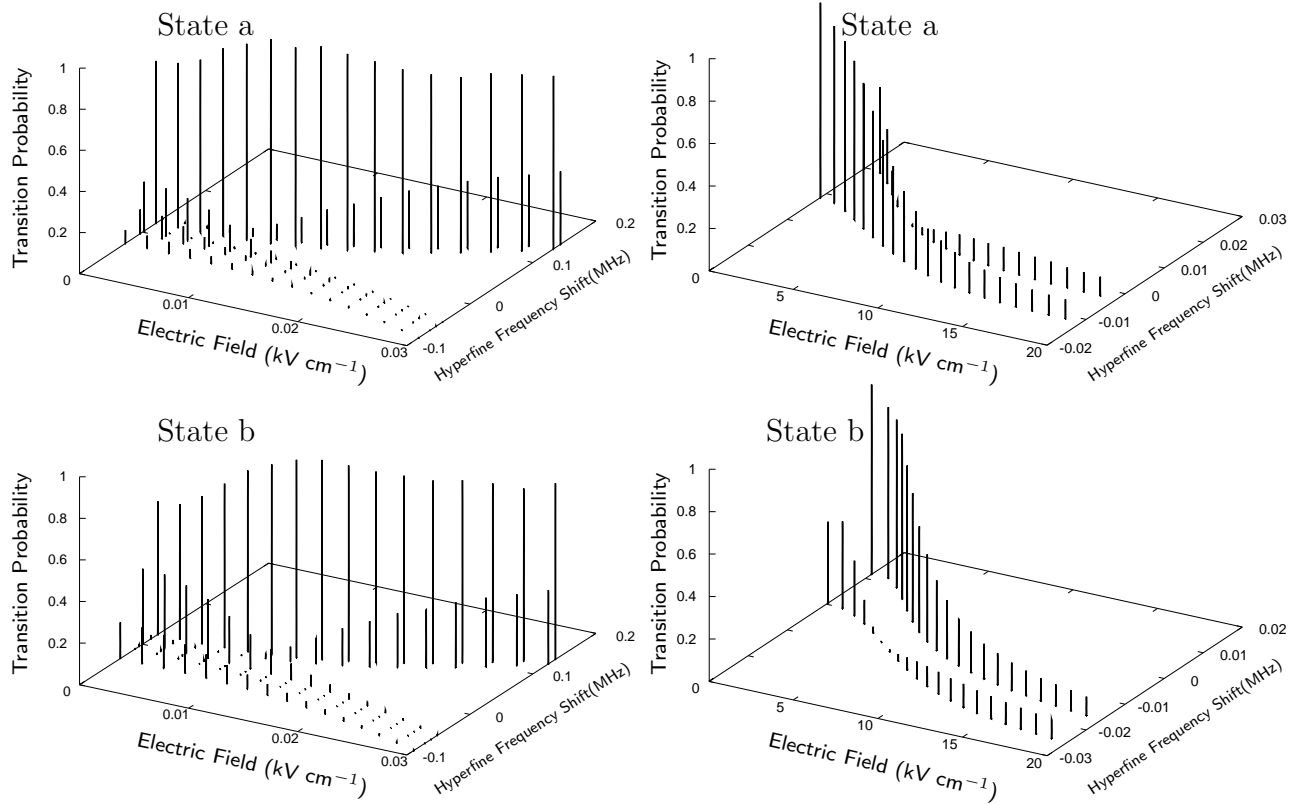}
\caption{\label{fig:10}%
Relative intensities for the transitions between the levels
correlating with $N$=0 (states a and b) and $N$=1 for
$^{7}$Li$^{133}$Cs ($M_{\rm{F}}$=+4). The microwave field is
polarized parallel to the electric field. The left-hand panels
show the spectra for low electric fields where $M_{\rm{N}}$ is
not yet a good quantum number. The right-hand panels show
spectra at higher fields where $M_{\rm{N}}$ is well defined and
only transitions to the ($N$=1, $M_{\rm{N}}$=0) branch are
permitted. In order to keep the spectrum on a single frequency
scale as a function of the electric field, we plot the
hyperfine frequency shift instead of the absolute frequency.
The most intense transition is assigned a peak intensity of 1.}
\end{figure*}

Close avoided crossings such as those shown in Fig.\
\ref{fig:08} might also cause nonadiabatic transitions between
hyperfine states for molecules in time-varying electric fields.
This might occur either when sweeping an electric field or for
molecules moving in an inhomogeneous field.

Figures \ref{fig:09} and \ref{fig:10} show the hyperfine
transitions driven by a $z$-polarized microwave field ($\Delta
M_{\rm{F}}$=0) between the $N$=0 (states a and b) and $N$=1
hyperfine states of $^{41}$K$^{87}$Rb ($M_{\rm{F}}$=+2) and
$^{7}$Li$^{133}$Cs ($M_{\rm{F}}$=+4) respectively. The
left-hand panels in these figures present blow-ups of the
weak-field regions of the spectra, where the separation between
the $|M_{\rm{N}}|$=0 and 1 levels takes place. At very low
fields $M_{\rm{N}}$ is not a good quantum number and the six
$N$=1 hyperfine levels can all be reached from both states a
and b. However, as the electric field increases $|M_{\rm{N}}|$
becomes better defined and the transition intensity
concentrates in the $M_{\rm{N}}$=0 branch (at higher
frequency). This happens at smaller electric fields for
$^{7}$Li$^{133}$Cs due to its larger electric dipole moment.

The right-hand panels of Figs. \ref{fig:09} and \ref{fig:10}
show the hyperfine structure of the spectra at larger fields,
where only transitions into the two ($N$=1, $M_{\rm{N}}$=0)
levels are permitted. These figures include the range of fields
where DeMille's quantum computing scheme would operate: 7.7 to
19.3 kV/cm for $^{41}$K$^{87}$Rb and 6.1 to 10.2 kV/cm for
$^{7}$Li$^{133}$Cs.

%
%
\begin{figure*}[t]
\includegraphics[width=0.95\linewidth]{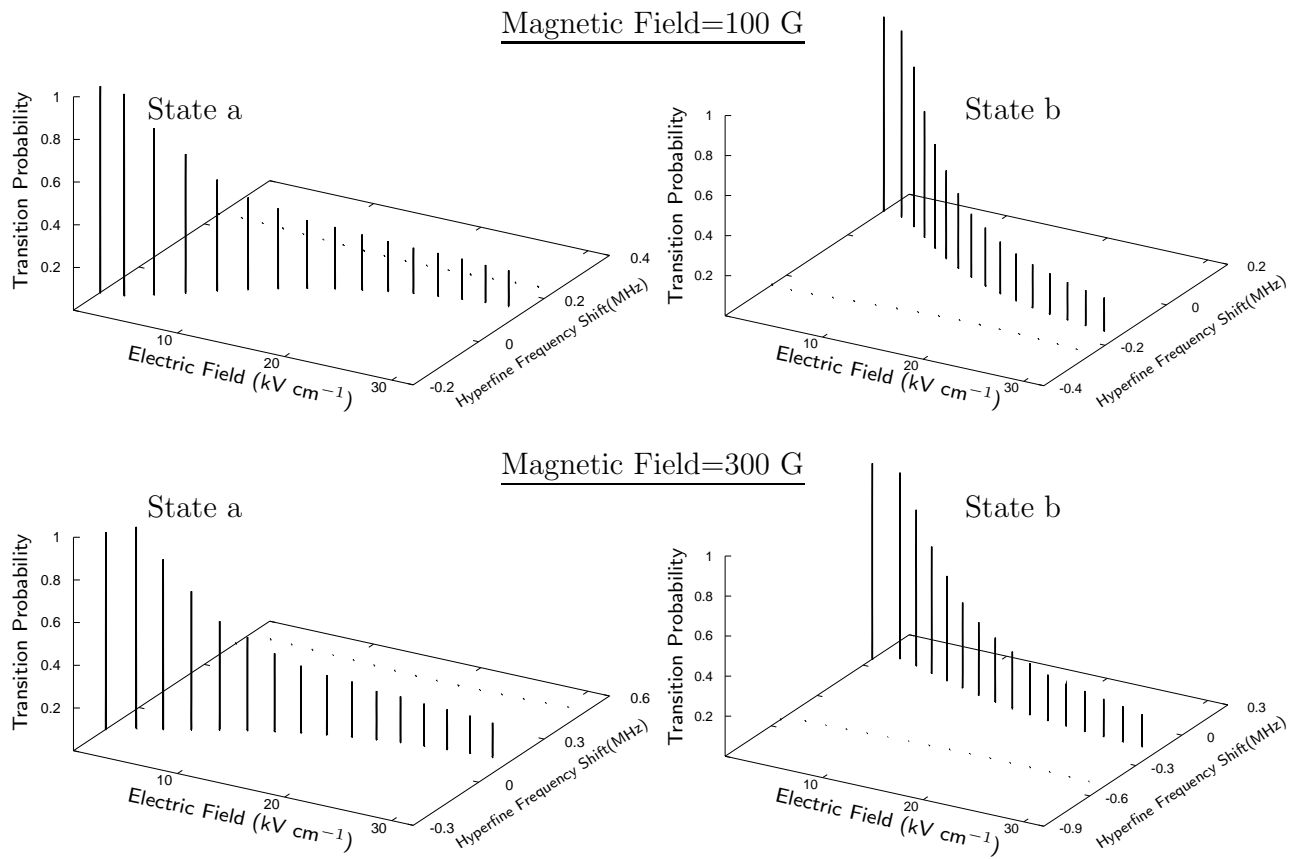}
\caption{\label{fig:11}%
Relative intensities for the transitions between the
$M_{\rm{N}}$=0 levels correlating with $N$=0 (states a and b)
and $N$=1 for $^{41}$K$^{87}$Rb ($M_{\rm{F}}$=+2) in parallel
electric and magnetic fields. The lowest electric field shown
is 0.3~kV~cm$^{-1}$. The microwave field is polarized parallel
to the external fields. In order to keep the spectrum on a
single frequency scale as a function of the electric field, we
plot the hyperfine frequency shift instead of the absolute
frequency. The most intense transition for each value of the
magnetic field is assigned a peak intensity of 1.}
\end{figure*}

%
%
\begin{figure*}[t]
\includegraphics[width=0.95\linewidth]{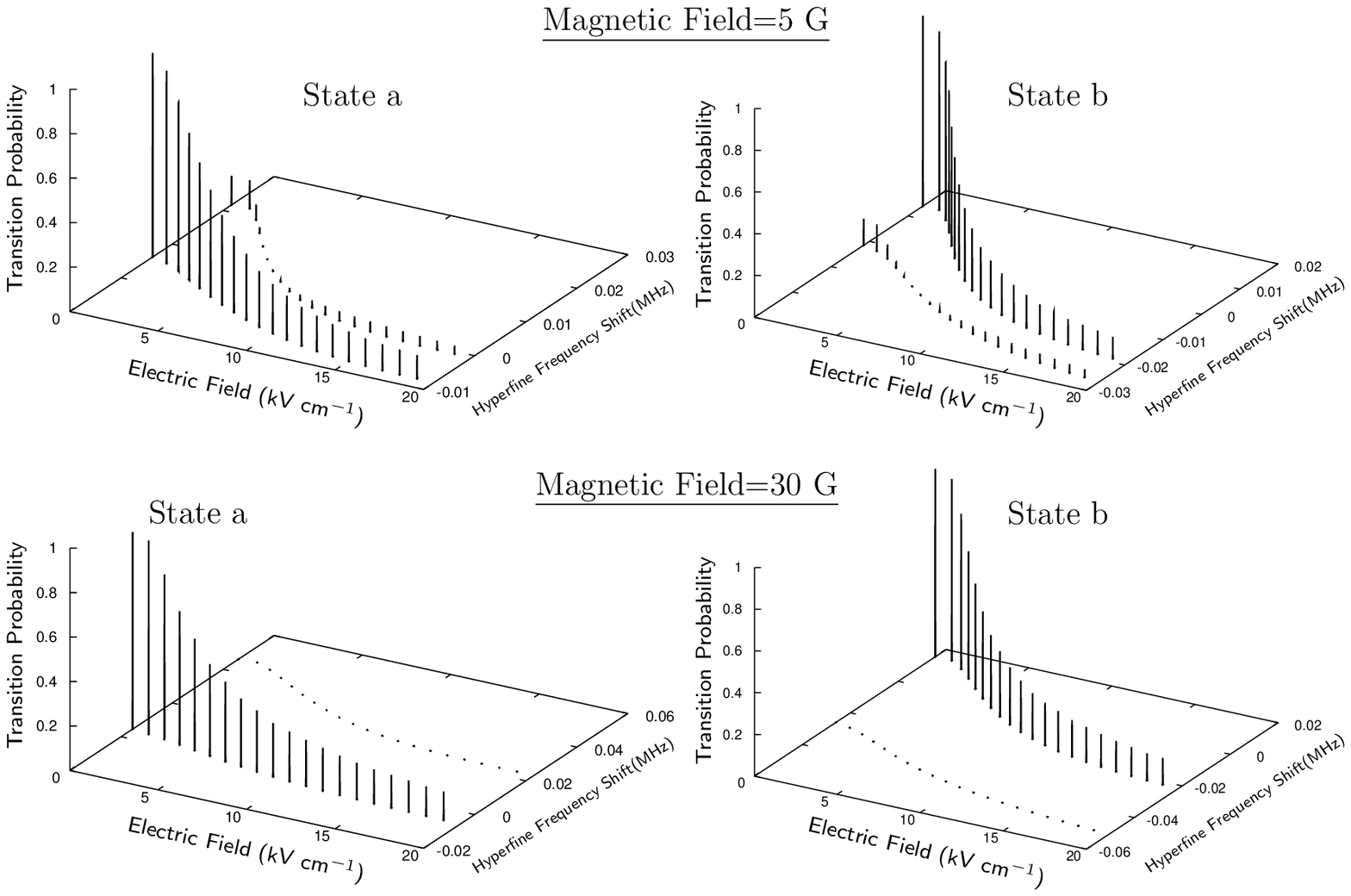}
\caption{\label{fig:12}%
Relative intensities for the transitions between the
$M_{\rm{N}}$=0 levels correlating with $N$=0 (states a and b)
and $N$=1 for $^{7}$Li$^{133}$Cs ($M_{\rm{F}}$=+4) in parallel
electric and magnetic fields. The lowest electric field shown
is 0.03~kV~cm$^{-1}$. The microwave field is polarized parallel
to the external fields. In order to keep the spectrum on a
single frequency scale as a function of the electric field, we
plot the hyperfine frequency shift instead of the absolute
frequency. The most intense transition for each value of the
magnetic field is assigned a peak intensity of 1.}
\end{figure*}

If the alkali dimers are going to work as qubits, it must be
possible to switch repeatedly between the $|0\rangle$ and
$|1\rangle$ states without exciting any other state of the
molecule and without altering the state of neighboring
molecules. The hyperfine structure will complicate the
operation by making these conditions more difficult to fulfil:
Figs.~\ref{fig:09} and \ref{fig:10} show that for both
molecules the spectra display several transitions with
significant intensity at each field in the working range. The
simulation of the spectrum for $^{41}$K$^{87}$Rb in the region
where the ($N$=0,$M_{\rm N}$=0) states cross, 18 to 20 kV/cm,
displays two lines with approximately the same intensity. The
insets of Fig.~\ref{fig:09} show a blowup of this region. The
hyperfine structure will also hinder the operation of the
device at electric fields below the crossings: the intensity of
the weaker transition is significant, mostly $\geq 10^{-4}$ of
that of the main peak, and the two lines spread over a range of
frequency shifts comparable to the frequency step used for
addressing (250 kHz). The spectrum for $^{7}$Li$^{133}$Cs
(Fig.\ \ref{fig:10}) has different features. It shows two
intense lines for the whole range of electric fields. Both
peaks are compressed in a 10 kHz interval of frequency shifts,
well below the addressing step for this molecule (1.7 MHz).

The different behavior in the relative intensities for
$^{41}$K$^{87}$Rb and $^{7}$Li$^{133}$Cs can be rationalized in
terms of the following rules (which apply only to
$M_{\rm{N}}$=0 states). For molecules where the nuclear
quadrupole interaction dominates the other hyperfine terms and
one of the nuclear quadrupole coupling constants is much larger
than the other, the individual nuclear spin projections $M_{1}$
and $M_{2}$ are nearly good quantum numbers for any electric
field large enough to make the $|M_{\rm{N}}|$$\ne$0
contributions negligible. The electric field not only causes
the state to be dominated by $M_{\rm{N}}$=0 but also indirectly
makes $M_{1}$ and $M_{2}$ well defined. This explains why for
$^{41}$K$^{87}$Rb one transition is much more intense than the
other except in the low-field and the crossing regions.
$^{40}$K$^{87}$Rb \cite{Aldegunde:spectra:2009} is similar in
this respect. By contrast, if the nuclear quadrupole
interaction is not dominant or if the two nuclear quadrupole
coupling constants are similar, $M_{1}$ and $M_{2}$ do not
become well defined as the field increases. In this case more
than one intense line will appear even at high fields, as in
the case of $^{7}$Li$^{133}$Cs.

In summary, the individual addressing of alkali dimers in a
quantum computing device such as that proposed by DeMille
\cite{DeMille:2002} may be hindered by the existence of
hyperfine structure. For molecules with large nuclear
quadrupole constants, the hyperfine spectra of adjacent
molecules may overlap, introducing an additional complication
to the operation of the device.

\section{Microwave spectrum in combined electric and magnetic fields}
\label{sec:hfBE}

Combining electric and magnetic fields provides an additional
degree of control over molecular properties and interactions.
In this section we explore the effects of applying parallel
electric and magnetic fields on the microwave spectra.

Figures \ref{fig:11} and \ref{fig:12} show the hyperfine
transitions between $N$=0 (states a and b) and $N$=1 levels of
$^{41}$K$^{87}$Rb ($M_{\rm{F}}$=+2, $M_{\rm{N}}$=0) and
$^{7}$Li$^{133}$Cs ($M_{\rm{F}}$=+4, $M_{\rm{N}}$=0) as a
function of electric field for different values of the magnetic
field. The two molecules show quite similar behavior. A
magnetic field resolves the near-degeneracy of the ($N$=1,
$M_{\rm{N}}$=0) states with different values of $M_{1}$ and
$M_{2}$. Once $M_{1}$ and $M_{2}$ become nearly good quantum
numbers, the spectrum is dominated by a single line. Even a
relatively small magnetic field is sufficient to achieve this
for both molecules, and there are no shallow avoided crossings
at high magnetic field such as those that appear for
$^{41}$K$^{87}$Rb in the absence of an electric field, as
discussed in Section \ref{sec:hfB}.

The subsidiary transitions are 6 or 7 orders of magnitude
weaker for $^{41}$K$^{87}$Rb at 500~G or $^{7}$Li$^{133}$Cs at
100 G and their intensities continue to decrease at higher
magnetic fields. Similar behavior is expected for all other
alkali metal dimers. It is therefore possible to produce a
single-line-dominated spectrum, with no crossings as a function
of electric field, simply by applying an additional magnetic
field. The use of combined electric and magnetic fields thus
restores the simplicity needed to apply DeMille's quantum
computing scheme.

\section{Conclusions}
\label{sec:con}

We have explored the hyperfine structure of the microwave
spectra of ultracold alkali metal dimers in applied magnetic,
electric and combined fields, taking full account of hyperfine
structure. We have compared the spectra for $^{41}$K$^{87}$Rb
and $^{7}$Li$^{133}$Cs, which have large and small values,
respectively, of the nuclear quadrupole coupling constants. Our
results can therefore be extrapolated to understand the
microwave spectra of any other alkali dimer.

The alkali dimers have very complicated spin structure arising
from the presence of two non-zero nuclear spins. Because of
this, there are many possible spectra that might be considered.
We have focussed mostly on spectra in which the initial
hyperfine state has $N$=0 and the same value of $M_{\rm F}$ as
Feshbach molecules produced from atoms in their absolute ground
states. However, our programs can readily be applied to other
$M_{\rm F}$ states and most of our conclusions can be extended
straightforwardly to such states.

The zero-field splitting is much larger for $N$=1 than for
$N$=0 because of the presence of nuclear quadrupole coupling.
At low magnetic fields and zero electric field, the microwave
spectra display many lines of significant intensity. This
persists until the nuclear Zeeman term dominates the zero-field
splitting for $N$=1. Some of the structure starts to disappear
at relatively low magnetic fields, of as little as a few Gauss,
but some persists to fields of thousands of Gauss in systems
with large quadrupole coupling constants where one of the
nuclear $g$-factors is small, such as in $^{41}$K$^{87}$Rb. At
relatively low fields, it is possible to find 2-photon paths
that allow molecules to be transferred between $N$=0 hyperfine
states via $N$=1 states. However, in the high-field limit the
spectra are single-line-dominated and no such paths exist.

An electric field mixes different rotational states and creates
orientation and alignment. The oriented states have possible
applications in quantum computing \cite{DeMille:2002}. For
oriented molecules, the levels correlating with $N$=0 and 1
both have substantial nuclear quadrupole structure. Even a very
small electric field is sufficient to separate the $N$=1 states
with $M_{\rm N}$=0 and $\pm1$ enough to prevent significant
mixing between them. This simplifies the spectrum to some
extent. However, the hyperfine levels for $N$=1 display close
avoided crossings as a function of field, which arise at points
where the oriented molecules have zero alignment. Even away
from the crossings, there is more than one transition with
significant intensity at each electric field, and this may
cause complications in designing a quantum computer.

Combined electric and magnetic fields provide a convenient way
to restore simplicity in the microwave spectra and eliminate
unwanted transitions. The magnetic field resolves the
near-degeneracy that exists when it is not present.

The calculations described in the present paper are also likely
to be important in other applications of ultracold polar
molecules, such as in the creation of novel quantum phases
\cite{Buechler:2007, Micheli:2007} and the development of
quantum simulators for condensed-phase problems
\cite{Micheli:2006, Wall:2009}. The spin structure must be
taken into account in a full treatment of collisions involving
alkali metal dimers.

\section*{Acknowledgments}
The authors are grateful to EPSRC for funding the collaborative
project QuDipMol under the ESF EUROCORES Programme EuroQUAM and
to the UK National Centre for Computational Chemistry Software
for computer facilities. HR is grateful to the China
Scholarship Council for funding her joint PhD student program
in Durham.


\end{document}